# Bridging the Gap: Unravelling Local Government Data Sharing Barriers in Estonia and Beyond

Katrin Rajamäe Soosaar[1] and Anastasija Nikiforova[2]

[1]Association of Estonian Cities and Rural Municipalities, Tartu, Estonia, katrin.raja@gmail.com

[2]University of Tartu, Tartu, Estonia, Nikiforova.anastasija@gmail.com

**Abstract**: Open Government Data (OGD) plays a crucial role in transforming smart cities into sustainable and intelligent entities by providing data for analytics, real-time monitoring, and informed decision-making. This data is increasingly used in urban digital twins, enhancing city management through stakeholder collaboration. However, local administrative data remains underutilized even in digitally advanced countries like Estonia. This study explores barriers preventing Estonian municipalities from sharing OGD, using a qualitative approach through interviews with Estonian municipalities and drawing on the OGD-adapted Innovation Resistance Theory model (IRT). Interviews with local government officials highlight ongoing issues in data provision and quality. By addressing overlooked weaknesses in the Estonian open data eco-system and providing actionable recommendations, this research contributes to a more resilient and sustainable open data eco-system. Additionally, by validating the OGD-adapted Innovation Resistance Theory model and proposing a revised version tailored for local government contexts, the study advances theoretical frameworks for understanding data sharing resistance. Ultimately, this study serves as a call to action for policymakers and practitioners to prioritize local OGD initiatives, ensuring the full utilization of OGD in smart city development.

## 1 Introduction

Open Government Data (OGD) plays a pivotal role in the development, with the OGD published by the local government contributing significantly to the development of smart cities and their transformation into more sustainable and smart entities. By providing data needed for analytics, real-time monitoring, and informed decision-making, OGD supports efficient urban planning and management (Reisi et al., 2020; Sharifi, 2020; Tsagkis et al., 2023; Lnenicka et al., 2023; Weil et al., 2023). Today, open data see an increased application in urban digital twins (or smart city digital twins), which further enhances cities' ability to plan, simulate, and manage urban environments efficiently and sustainably (Adreani et al., 2024; Conde et al., 2023; Moghadas et al., 2022). This is often carried out in collaboration with stakeholders through co-design and co-creation mechanisms and as such open data is often referred to as „urban innovation" (de Magalhães Santos & Madaleno, 2023). As such, OGD facilitates the maturation of both urban environments and society as an integral part of the open data ecosystem and smart city ecosystem (Lnenicka et al., 2023), advancing the development of sustainable and smart cities and societies.

Despite its potential, the full value of local administrative data— closest to cities citizens often generated by citizens, industries, academia, and public authorities—remains underutilized due to organizational, technological, and legal barriers (Liva et al., 2023; Fusi, 2021; Sandoval-Almazan et al., 2021). These challenges are prevalent in both underdeveloped and technologically advanced countries. For instance, Estonia, known as a "digital state," has faced difficulties in providing and maintaining OGD despite its success in digital governance (McBride et al., 2018). Significant progress has been made since 2020, with Estonia climbing from 24th place in 2018 to 5th place in the European Open Data Maturity Report and maintaining a top position due to improvements in policy indicators and impact dimensions. However, persistent challenges in data provision and quality at the local government level remain (Rajamäe-Soosaar & Nikiforova,

2024). As of mid-2023, only 12 out of 79 (15%) Estonian local governments (15 urban municipalities and 64 rural municipalities), had published any data on the national OGD portal.

A study conducted in 2022, examining the information systems of local government revealed that the information systems of municipalities in Estonia do not form a coherent integrated ecosystem (Proud Engineers, 2024). Knowledge of service management and development is generally low, and most local governments do not prioritize data management and open data. The Estonian Digital Agenda 2030 also highlights the digital divide between local authorities and state agencies. This underscoring the need for a more in-depth examination of municipal issues to investigate possible causes.

The objective of this study is to examine Estonian OGD development with a particular focus on the local level, identifying the main barriers municipalities face when openly sharing OGD. To achieve the objective, we conduct qualitative analysis of barriers faced by local governments when opening their data through interviews with local governments, whose protocol is developed by drawing from the OGD-adapted Innovation Resistance Theory (IRT), which is intended to study functional and psychological factors OGD barriers, empirically identifying predictors affecting public agencies' resistance to openly sharing government data (Nikiforova and Zuiderwijk, 2022). While focusing on examining barriers associated with the governmental data sharing by local governments, this study also validates the model, as it remains so far to be conceptual model with no application to the real-world context.

The findings from local government interviews are used to propose improvements for the current Estonian OGD ecosystem and its local data ecosystem in particular. As one of the authors is involved with the Estonian Association of Municipalities, which represents the common interests and arranges co-operation of local governments being responsible for developing and implementing the digital transformation plan of local governments, we intend to use these results to support municipalities in developing data governance and open data initiatives by transforming these recommendations into actions.

Given the limited research on local OGD or local data ecosystem, this study offers new insights into the role of local governments/ municipalities in providing open data and the associated barriers and enablers. By focusing on Estonia, the study serves as call for action for more research and practical efforts to address challenges at the local level (urban data ecosystem), as the integral yet often overlooked part of open data ecosystem, aiming for a more resilient and sustainable open data ecosystem and smart city (data) ecosystem in particular. This is even more so, considering that we demonstrate that even in the case of very successful national OGD initiative (as evidenced by high ranking in the OGD indices and benchmarks), it may lag at the local level. This discrepancy highlights the need for focused attention on enhancing OGD efforts within municipalities to match the nation's trend-setting status. The results could inform data governance and open data policymaking, offering practical recommendations for national and local authorities to enhance Estonia's open data landscape.

The paper is structured as follows: Section 2 provides background on the topic, Section 3 presents methodology. Section 4 presents the results of the interviews with local governments' representatives on barriers and enablers towards OGD publishing and maintenance and the recommended measures for local data ecosystem improvement. Section 5 establishes discussion, acknowledges limitations, suggests future research directions, while Section 6 concludes the paper and stresses practical and theoretical contributions of this study.

## 2 Background

### 2.1 Related literature

The OGD ecosystem at local and regional levels in Europe shows slow progress. This is particularly true for Estonia, despite its advanced e-government infrastructure, which has rarely been the focus of OGD research, with only a few studies investigating the reasons for its low OGD maturity, although both at the national level (Kassen, 2019; McBride et al., 2018). There has been no recent examination of Estonia's rapid evolution from "follower" to "trend-setter" (in ODM reports). Research on local government-level OGD in Estonia is notably lacking, a trend seen across Europe. Most studies at European level that focus on the local data ecosystem, concentrate on open data platforms, data usage and sharing, and digital maturity assessments in municipalities, with limited attention to local OGD and barriers public agencies face when opening data, as evidenced by a systematic literature review by Rajamäe-Soosaar & Nikiforova (2024).

Given the potential of city and municipal data, there is a need for research on regional and local OGD ecosystems, focusing on the progress of open data initiatives, barriers to data sharing, and user perspectives. While OGD adoption has been extensively studied at the national level, the used methods can be adapted to the local government context, e.g., established technology adoption models. However, current research primarily emphasizes technology acceptance models, which aim to identify factors predicting technology use, while overlooking the reluctance towards data opening.

In addition, most studies primarily investigate OGD acceptance from the user perspective (e.g., Wirtz et al. (2019), Talukder et al. (2019), Zuiderwijk et al. (2015), Shao (2023), Wang et al., 2020; Islam et al., 2023). A few studies explore the OGD provider's perspective, all elucidating factors that impact adoption by government agencies. For instance, Wang and Lo (2016) used the Technology–Organization–Environment (TOE) framework to explore key OGD adoption determinants in Taiwan. Hossain et al. (2022) proposed a model integrating seven context-specific TOE variables to understand factors shaping OGD initiatives. Huang and Huang (2022) examined individual and organizational factors influencing public servants' intention to implement OGD policies in Taiwan, using the Theory of Planned Behaviour (TPB). As such, there is a lack of research on the resistance towards OGD adoption with the use of theoretical models. The reluctance perspective with identification of surrounding barriers towards data opening, remains crucial as being capable to pinpoint actual issues and challenges that need to overcome to success-fully adopt the phenomenon. The Innovation Resistance Theory (IRT) offers a promising approach to identifying barriers preventing the acceptance of OGD. To this end, OGD-adapted model has been proposed by Nikiforova and Zuiderwijk (2022), though it has never been tested, i.e., it remains a theoretical concept that requires validation through its application in real-world scenarios with its further refinement, should such be needed. To this end, this study addresses this gap by using the proposed theoretical model to examine barriers Estonian local governments are facing, while by doing so, validating the model and evaluating its general appropriateness to the studied problem.

## 2.2 Estonian OGD journey

Estonia launched its first national open data portal in 2015. Initially, it hosted a small number of datasets, where by 2018 – 3 years after the portal launch - data availability was limited to 89 datasets, found by users inconvenient to use it (McBride et al., 2018; McBride et al., 2020). A major factor affecting insufficient focus on open data movement in Estonia is typically considered to be due to the X-Road - the national interoperability infrastructure that provide a unified and secure data exchange system, used by Estonian public sector institutions for exchanging data and offering services to the citizens, thereby reducing the emphasis on open data initiative and making policymakers sceptical regarding OGD initiatives (McBride et al., 2018). The state of the art, however, changed with the societal and international pressure, including the importance of OGD in international e-Government rankings (McBride et al., 2020), and entering into a contract with Open Knowledge Estonia (non-profit organization) to improve the performance of Estonia's OGD ecosystem done by the Estonian Ministry of Economic Affairs and Communications in 2018, which led to substantial improvements. By 2020, the number of datasets, unique users, applications, and open data events had significantly increased, propelling Estonia's ranking in the European Open Data Maturity report from 27th to 14th place (McBride et al., 2020). A pivotal role in this transformation was played by legislation.

By 2021, amendments to the Estonian Public Information Act aligned with EU Directive 2019/1024, imposing new obligations for public information provision and re-use. These included requirements for open data descriptions, availability of research data, immediate release of dynamic data, non-restrictive re-use conditions, and free access to high-value datasets.

As a result of this transformation, Estonian efforts have been recognized in various EU and global rankings. E.g., the European Open Data Maturity Report shows Estonia's impressive journey from 27th place in 2018 to a "trendsetter" position by 2020, reaching 4th place in 2023. This progress highlighted substantial improvements in policy indicators and the impact dimension, despite persistent challenges in data provision from smaller local governments and issues with the quality of open data, particularly regarding the currency of data and metadata, and linked data. In the OECD OURdata Index of 2023, Estonia ranked 4th out of 36 countries, excelling in policy content and government data literacy programs. However, the country scored lower in the availability of high-value datasets and impact monitoring. The most recent report by the Open Data Inventory (ODIN) ranked Estonia 5th in Northern Europe and the 11th position globally (as of 2022). ODIN also identified coverage as a key area for improvement, particularly in social statistics and subnational data availability, based on data from Statistics Estonia. This again points to the shortcomings in local-level data provision. The World Justice Project, while not primarily focused on open data governance, ranks Estonia 9th globally and 7th regionally in 2023. Estonia scores highly for publicized laws and government data but ranks lower in "Civic Participation" and "Complaint Mechanisms" (13th out of 31 countries). Overall, Estonia's journey in open data governance showcases significant progress and leadership, especially at the national level, while highlighting ongoing challenges in local data provision and quality.

Most benchmarks focus primarily on central governments, where the Global Open Data Index is one of those few that covers regional or local levels, but it has not covered Estonia in any of its assessments. Unfortunately, this project has been archived, limiting in-depth insights into local open data landscape.

In 2021, the new open data portal (avaandmed.eesti.ee) was launched, which is powered by a custom-made platform for the open data portal tailored to meet specific needs better than the CKAN platform that powered the first Estonian OGD portal. By January 2024, the portal hosted 1,807 datasets, reflecting significant growth. However, only 12 out of 79 local governments in Estonia had contributed to the portal, publishing fewer than 170 datasets in total. The primary contributors among these were the municipalities of Tallinn and Tartu. Both cities have also dedicated portals for geospatial data.

Low local governments participation rate, however, may be due to the fact that, in accordance with the Estonian Public Information Act, local governments must maintain a document register and disclose certain documents, such as legislation, contracts, and public letters, from their Document Management System (DMS). The public document register includes documents not subject to restrictions, such as private data. Municipal websites typically feature direct links to the public view of their DMS, allowing access to JSON files containing legislation, contracts, and official correspondence from nearly 50 local governments. Few municipalities have published links to their DMS datasets on the OGD portal. For Anija, Harku, and Saue Municipalities, the portal notes that the data was migrated from the previous version of the national OGD portal without publisher review. Nine other local governments have updated their dataset metadata on the national portal. Most municipalities, however, have not done none of the above providing the access to their DMS through their websites. To this end, as we found out during the discussion with the representative of the Ministry of Economic Affairs and Communications, a new Data Portal/Gateway will be developed soon, integrating the State Information System management system (RIHA) and the Open Data Portal, with improved usability for a better user experience. To address these challenges, Estonia plans to integrate the State Information System management system (RIHA) with the Open Data Portal and improve its usability. However, despite high rankings and increase in popularity of the OGD portal, indicated by the number of datasets it hosts and usage statistics (download rates, rating and feedback along with the use cases (134 use-cases as of May 2024)), within the local context of Estonia open government data (OGD) is not widely recognized. A 2024 survey[1] for the Ministry of Economic Affairs and Communications found that only one-third of Estonians had heard of "open data," and just 17% understood its meaning. However, 71% of respondents still considered open data important for society. In summary, while Estonia has improved its positions in many open data rankings, local governments are still lagging, as evidenced by the modest number of municipalities publishing data on the national portal. A survey on open data awareness revealed that the topic is relatively unknown to the public, which may also be a constraint at the local administrative level. These findings are used as for developing an interview protocol to identify the main barriers local governments face in sharing and maintaining open data, directly addressing the research question of this study.

## 3 Methodology

To answer the research question, we conduct a qualitative study driven by technology adoption theory to empirically identify the main barriers Estonian local governments face in sharing and maintaining open data.

To this end, we utilize the model proposed by Nikiforova and Zuiderwijk (2022) to study barriers associated with the OGD opening. They propose an OGD-adapted IRT model to study the resistance of public authorities to openly share government data.

Innovation Resistance Theory (IRT), in turn, was developed by Ram and Sheth in 1989 (Ram and Sheth, 1989) arguing that consumers resist innovations despite their necessity and desirability due to several barriers that inhibit their adoption. According to IRT, barriers are categorized into: (1) functional barriers that arise when consumers face significant changes from adopting the innovation, related to product usage, value, and risks; (2) psychological barriers, which stem from consumers' traditions and perceived product image, often conflicting with prior beliefs.

IRT has been widely applied in marketing and business research (Ma and Lee, 2019; Dwivedi et al., 2023), with many studies empirically evaluating consumer resistance to innovations (Talwar et al., 2020). However, its application in the e-government domain is limited to two studies, with (Prakash and Das, 2022) that examines citizens' resistance to digital contact tracing apps during the COVID-19 pandemic and (Nikiforova & Zuiderwijk, 2021) that proposed the model we are using in this study.

Nikiforova & Zuiderwijk's (2021) conceptual model consists of five main IRT barrier categories, each associated with OGD barriers identified through a literature review (Table 1). Based on these barriers, five hypotheses were formulated to study public authorities' resistance to openly sharing government data in the form of *"[Construct ∈ {Usage barrier; Value Barrier; Risk barrier; Tradition Barrier; Image Barrier}] has a positive effect on public agencies' resistance toward openly sharing government data"* (H1: Usage barrier, H2: Value Barrier, H3: Risk barrier, H4: Tradition Barrier, H5: Image Barrier) (Nikiforova & Zuiderwijk (2021, p.3). We employ these hypotheses, tailoring them to the specific context of our study. As such, our hypotheses are:

> **H1:** *Usage barrier (UB) positively influences local governments' resistance to openly sharing government data .*
> 
> **H2:** *Value barrier (VB) positively influences local governments' resistance to openly sharing government data .*
> 
> **H3:** *Risk barrier (RB) positively influences local governments' resistance to openly sharing government data .*
> 
> **H4:** *Tradition barrier (TB) positively influences local governments' resistance to openly sharing government data .*

---

[1] Open data portal (eesti.ee)

**H5:** *Image barrier (IB) positively influences local governments' resistance to openly sharing government data.*

Table 1. The proposed model and its elements (Nikiforova & Zuiderwijk, 2022)

| Barrier | Measurement item |
|---|---|
| Usage Barrier (UB) | UB1: It is difficult to attain the appropriate quality level for open government data to be shared openly <br> UB2: It is difficult to prepare data for publication so that they comply with OGD principles <br> UB3: It is difficult to prepare data for publication so that they become appropriate for reuse <br> UB4: Data are difficult to publish on the OGD portal due to the complexity of the process <br> UB5: Data are difficult to publish on the OGD portal due to the unclear process <br> UB6: Data are difficult to publish on the OGD portal due to their limited functionality <br> UB7: Open government data portals often do not allow for semi-automation of the publishing process <br> UB8: It is difficult to maintain openly shared government data |
| Value Barrier (VB) | VB1: My organization believes that openly sharing government data is often not valuable for the public <br> VB2: Many open government datasets are not appropriate for reuse <br> VB3: Many open government datasets suffer from data quality issues (completeness, accuracy, uniqueness, consistency etc.) <br> VB4: The public gains of openly sharing government data are often lower than the costs <br> VB5: My organizations' gains of openly sharing government data are often lower than the costs <br> VB6: Data preparation is too resource-consuming for my organization <br> VB7: Open government data do not provide any value to my organization <br> VB8: Open data that my organization can openly share will not provide value to users <br> VB9: The amount of resources to be spent to prepare, publish and maintain open government data outweigh the benefit my organization gains from it |
| Risk Barrier (RB) | RB1: My organization fears the misuse of openly shared government data <br> RB2: My organization fears the misinterpretation of openly shared government data <br> RB3: My organization fears that openly shared government data will not be reused <br> RB4: My organization fears violating data protection legislation when openly sharing government data <br> RB5: My organization fears that sensitive data will be exposed as a result of opening its data <br> RB6: My organization fears making mistakes when preparing data for publication <br> RB7: My organization fears that users will find existing errors in the data <br> RB8: My organization fears that openly sharing its data will reduce its gains (otherwise the organization could sell the data or use it in another beneficial way) <br> RB9: My organization fears that openly sharing its data will allow its competitors to benefit from this data |
| Tradition Barrier (TB) | TB1: Freedom of information requests are sufficient for the public to obtain government data <br> TB2: My organization is reluctant to implement the culture change required for openly sharing government data <br> TB3: Employees in my organization lack the skills required for openly sharing government data <br> TB4: Employees in my organization lack the skills required for maintaining openly shared government data <br> TB5: My organization is reluctant to radically change the organizational processes that would enable openly sharing government data |
| Image Barrier (IB) | IB1: My organization has a negative image of open government data <br> IB2: My organization believes that open government data is not valuable for users <br> IB3: My organization fears that openly sharing government data will damage the reputation of my organization <br> IB4: My organization fears that the accidental publication of low-quality data will damage the reputation of my organization <br> IB5: My organization fears that associating them to incorrect conclusions drawn from OGD analysis by OGD users will damage the reputation of my organization |

Figure 2 presents the model, with the numbers in brackets indicating the number of measurement items defined for each barrier, as presented in Table 1.

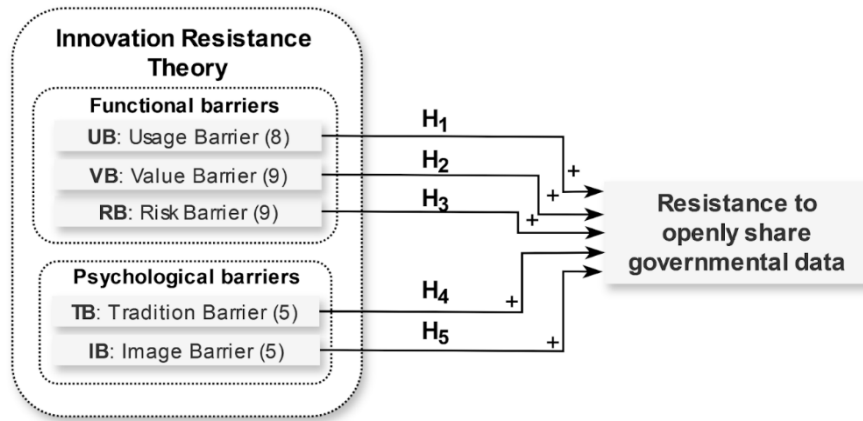

Figure 2. Research model and proposed hypotheses (the numbers in brackets refer to the number of measurement items in Table 1) (Nikiforova & Zuiderwijk, 2022)

While the model is approach-agnostic, allowing to utilize both the quantitative and qualitative approach, the authors recommend a qualitative study for richer insights, particularly in understanding the nuanced challenges of individual organizations, navigating conversation, considering the unique setting of an individual organization. As such, the study proposes conducting interviews with representatives of public agencies.

To this end, an interview protocol is developed based on the model, as described in the next section along with the sampling approach utilized in this study.

### 3.1 Interview protocol

The initial interview protocol, based on the proposed OGD-adapted IRT model by Nikiforova & Zuiderwijk (2021), was tailored to suit the objectives of this study. It was expanded to encompass (1) the OGD ecosystem (5 questions), and (2) organizations' experience with the Estonian national portal avaandmed.eesti.ee. As the interviews focused on local governments, all the questions were adapted to municipalities' perspective. The final interview protocol is available in Supplementary Materials and on ZENODO [link to be added upon acceptance]. To avoid language barrier, the interview protocol was translated into the respondent's native language – Estonian.

The interview protocol comprises four major sections. The first section delves into the general profile of the organization represented by the respondent and their awareness of open data. This includes inquiries about the local government, the respondent's role, the type of data collected by the municipality, and their awareness of the concept of "open data." The later question was added in the light of the survey that revealed limited awareness of open data amongst Estonian residents[2].

The second section addresses openly sharing data on municipal websites or portals and the national open data portal. Most of the Estonian local governments have not published data in national portal, but there might be datasets that are openly shared on the municipal website (e.g., document registers) (as found in (Soosaar & Nikiforova, 2024)). The Public Information Act indicates the data that should be disclosed in open format also by local governments, if possible and appropriate, and would not involve disproportionately great effort. Hence, this interview section was split into two separate blocks repeating questions to gain more insight on the experience with both hubs and identify potential datasets that possess the characteristics of open data on local level websites.

In the light of the above, this section contains questions:

1) whether the municipality has ever openly shared its own data or the data it collected from other sources;
2) what were the drivers for doing this or what were the reasons for not sharing them openly;

---
[2] https://avaandmed.eesti.ee/

3) *what type of data the municipality shared openly? and how often;*
4) *what was the process of openly sharing data? and who within the organization were involved;*
5) *what challenges the organization faced in openly sharing data;*
6) *what have been the specific cases, where municipality's open data was reused by third party?*

The latter question was added to this section to identify possible use cases of OGD usage, which could contribute to better understanding of the local data ecosystem. If the local government had not shared data on national open data, additional question about the reasons for not doing it were asked.

The third section focuses on IRT model-related questions. As this study is explorative in nature, each measurement item was converted into an open-ended question of the form *"To what extent do the following situations form a barrier to openly sharing your organization's data: ..."* where each barrier listed in Table 1 is then addressed.

Open-ended question and long responses were given preference over closed-ended questions (yes/no) as the interest was to delve into the respondents' real experiences with obstacles encountered by local governments. Before addressing specific barriers, respondents were provided with a definition of each barrier and asked if any challenges related to openly sharing data were encountered. The same question was asked after addressing each specific barrier, i.e., *whether there are any other barriers that could form a barrier?* (see Table A.1 1). After all the barriers' related questions have been asked, additional general question was included to identify any additional barriers not covered in the predefined categories. These additional questions would help to verify whether all potentially relevant barriers were captured, aiding in the validation and refinement of the model.

The final section explores the overall open data ecosystem and potential improvements. It seeks the respondent's perspective on various aspects, including their perception of the OGD ecosystem, awareness of policy documents or initiatives, suggested actions at national and municipal levels, and the perceived value of open data. As such, it consists of 5 questions:

1) *how the respondent sees the OGD ecosystem and its actors;*
2) *whether respondent is aware of policy documents or initiatives guiding developments of OGD on Estonian or European level;*
3) *which actions should be taken at the national level;*
4) *which actions by municipalities to improve the disclosure of open data by local governments;*
5) *what is the benefit or value of open data?*

## 3.2 Interview sample

When assembling the interview sample, selection criteria were established based on several factors for municipalities, namely (1) population size, (2) geographical diversity, (3) presence on the national open data portal.

Considering the wide variation in population across local governments, ranging from 165 registered inhabitants to over 50,000 residents with the average number of residents of 17 381 (According to Estonian Population Register[3] as of 01.01.2024), the selection aimed to encompass municipalities from three categories based on population size: those with fewer than 10,000 residents, those with 10,000 to 50,000 residents, and those with over 50,000 residents. To ensure representation from different regions, municipalities from each geographical area (e.g., Northern, Southern) were included in the sample. Finally, the local governments that have published data in national open data portal were identified based on results of the respective analysis in (Rajamäe-Soosaar and Nikiforova, 2024). The objective was to include approximately one-third of the municipalities already sharing data on the national portal, thereby incorporating their experience into the study. On the other hand, municipalities without prior experience with the national portal provided valuable insights into the barriers preventing data publication.

---
[3] https://www.siseministeerium.ee/en/activities/population-procedures/population-register

The selection process resulted in 16 local governments, out of which consented to participate in the interviews, forming the final sample. Informed consent for recording the interviews and utilizing the collected data was obtained from participants prior to the interviews, with both parties digitally signing the consent form.

All interviews were conducted online in Estonian using MS Teams software and were recorded. The recordings were then converted into MP3 audio files using VLC software and transcribed using an Estonian speech recognition and transcription tool Textiks[4]. The transcriptions were subsequently analysed using qualitative data analysis software NVivo, with coding based on the interview protocol questions. The list of codes, along with the number of files and references, is provided in the Table B.1, while the results are presented in the next Section.

# 4 Results of interviews with local governments on barriers and enablers towards OGD publishing and maintenance

This section presents results of interviews conducted with 12 local governments across Estonia, aimed at identifying the obstacles associated with the publication and maintenance of OGD at the local level in Estonia. Organized in alignment with the sections outlined in the interview protocol (as detailed in Section 2.2.1), the results provide insights into: (1) the general profile of the respondents, (2) the organization and its stance on open data, (3) usage barriers, (4) value barriers, (5) risk barriers, (6) tradition barriers, (7) image barriers, and (8) the broader landscape of the open data ecosystem. Finally, the most relevant barriers and the recommendations for actions at both the national and local levels are provided.

## 4.1 General profile of the respondents

Interviews with the 12 municipalities included in the sample were conducted between March and April 2024. In two cases, there were 2 respondents from one municipality, while the other 10 interviews were conducted one-on-one, involving a total of 14 respondents.

Half of the interviewed local governments have populations ranging from 10,000 to 50,000 inhabitants, while 3 municipalities have over 50,000 residents and 3 have fewer than 10,000 residents according to the population registry. Half of the respondents held top-level managerial positions such as Mayor, Deputy Mayor, or Member of Municipal Council. 29% were Top/Leading Specialists (e.g., Data Manager, IT Advisor), and 3 respondents held mid-level management positions (e.g., head of department).

## 4.2 Organization and provision of open data

In the first part of the interview exploration centred on the type of data collected and published by the organization represented by the interviewee, along with the associated challenges and drivers for openly sharing data.

Two municipalities (LG[5]4, LG7) underscored that all functions and tasks outlined in the Local Government Organisation Act[6] and other legislation necessitate data collection or processing. The types of data mentioned by all respondents include:

- data collected by providing (e-)services (e.g., benefits, permits, applications);

---
[4] tekstiks.ee
[5] Local Government
[6] https://www.riigiteataja.ee/en/eli/501072023003/consolide

- document management system (local governments regulations, correspondence, etc.);
- data about the municipality's personnel (e.g., salary data, interview summaries);
- data collected into national registers (e.g., population register, education information system);
- data collected due to legal obligation (e.g., register of pets, register of cemeteries);
- geospatial data connected to land use and infrastructure (e.g., detailed plans, lighting, playgrounds);
- surveillance and sensor data (e.g., cameras, pollution sensors);
- analyses and surveys (e.g., satisfaction surveys).

Some local governments (LG2, LG4, LG7) emphasized that while it is not mandatory to use GIS software and create WMS services of their spatial data, they have allocated additional financial and human resources to it. Two respondents highlighted that they collect additional data besides education information register, such as information about hobby/recreational education. One respondent (LG11) categorized the bases for data collection into three categories: data collected by law, based on consent, and data collected under contract.

Respondents' awareness of the concept of "open data" was examined to understand their prior exposure to this topic and put the focus on open government data. Most of the **respondents were aware of the term "open data"** and could mention several principles of open data, but many acknowledged that it is somewhat challenging to distinguish between "data" and "open data" in the context of local government. **Two-thirds of the interviewed municipalities were unaware neither of both general open data principles[7] and the obligation arising from the Public Information Act**, which stipulates that open data should, if possible, be published in machine-readable format and should include data descriptions describing datasets and data (metadata).

Questions about sharing organizational data on the municipal portal and national portal (section two of the interview) were divided into two separate blocks, as 7 of the interviewed local governments had not published any data on the national portal. The first part focused on **municipal website or portal**. Most respondents mentioned that they comply with § 28 of Public Information Act[8] that stipulates the obligations of local government agencies to disclose data and information on their website, but in addition, they publish more data, e.g., geospatial data. Table 2 summarizes the **main types of data that respondents openly share on their websites**. The format of these data varies greatly, and in many cases, machine-readable formats are not provided.

Table 2. Main type of data on municipal website/portal

| Type of data | Description | Cases |
| --- | --- | --- |
| Development plan, budget strategy and annual budget | The drafts and final versions of these documents should be available | All |
| Spatial plans | Comprehensive spatial plan and detailed spatial plans in line with Planning Act[9] | All |
| Services of municipality | Description of services, grants, permits etc. that the local authority offers to residents | All |
| Other geospatial data | Map layers, e.g., public transport routs and stops, local authority agencies, parking areas, public waste stations and containers, playgrounds etc. | LG1, LG2, LG11, LG4, LG5, LG7, LG9, LG 12 |
| Statistical data | Different statistics about the municipality | LG4, LG5, LG11, LG7, LG10 |
| Document register | Local government legislation, contracts, and public correspondence | All |
| Visualised budget implementation | Interactive visualisation of local government budget implementation by MS PowerBI | LG1, LG 10, LG3, LG 9 |
| Procurement plan | Annual plan of procurements | All |

---

[7] https://public.resource.org/8_principles.html
[8] https://www.riigiteataja.ee/en/eli/503052023003/consolide
[9] https://www.riigiteataja.ee/en/eli/504072023008/consolide

| | | |
|---|---|---|
| Contacts | Contacts of local authority bodies (council and administration) and agencies (e.g., schools, kindergartens, libraries) | All |

The **process of openly sharing data** on municipal websites varies depending on the size and internal processes of the local government. In some cases, the responsibility for publishing specific data on the municipal website or designated geoportal lies with the data owner or department. Smaller municipalities often have only one website administrator, who collects data from data owners or to whom the data owners turn when data needs to be updated.

The **primary driver for openly sharing data on municipal websites** (stimuli) in most cases is the obligation outlined in the Public Information Act and other legislation regulating the disclosure of specific information. However, some respondents emphasized that this is not the sole motive for public disclosure of local government data.

Other motives highlighted by the respondents include:

- transparency of the local government agencies (LG3, LG5, LG6, LG8);
- output, whose production was supported by public funding and therefore should be publicly available (driven by their own intent rather than law or regulation) (LG10, LG4);
- informing citizens and facilitating citizen autonomy (LG4, LG9, LG10).

Additionally, LG7 and LG9 pointed out that the data was initially collected to optimize their internal processes or service delivery, and afterward, it was openly shared with the public.

Interviewed organizations acknowledged various **challenges they have faced or are facing in openly sharing data on municipal websites**, which are categorized with examples in Table 3.

Table 3. Challenges that municipalities have faced in openly sharing data

| Challenge | Description or example of the challenge |
|---|---|
| Lack of human resources | Lack of human resources or current personnel is too engaged with day-to-day activities (LG6, LG8, LG10)<br>LG10: *"Each department is responsible for their own things on the website and if they don't have the skills to put it there or the rights, then the Communication Specialist has been there for that. Of course, when people have changed or someone is on vacation, a challenge always arises"* |
| Technical platform | The sharing of data should be done with less effort - technical platforms have limited functionalities (LG11, LG12) |
| Lack of competence and understanding | Lack of data competencies in the organization – data governance or management, data analytics, open data, data quality management etc. (LG3, LG5, LG1, LG9, LG10) |
| Security and privacy issues | Decision-making on data that can be publicly shared due to possible privacy breach. E.g., some spatial data (e.g., locations of sensors) could be reused for criminal motives (e.g., cyber-attacks, theft) or should be not published for national security reasons (LG1, LG2, LG9) |
| Reusability | Difficulties in data preparation to be shared as open data (open data principles-compliant) due to format in which they are stored by the organization (LG12)<br>LG12: *"There is a lot of information on the website that could actually be grouped under open data, but it's not in machine-readable format today."*<br>Possible mistakes when anonymizing data for sharing (LG9) |
| Organization processes | Difficult to implement the data management and publication process, so that every data owner understands and fulfils it (LG7) |

Local governments are cognizant of various stakeholders/ **parties who reuse the data** municipality shares, including:

1) students, who use the data for research purposes and projects, such as developing new applications;
2) journalists, who use financial data and document register as the basis for their articles;
3) actors involved in spatial planning (e.g., real estate developers), who use GIS portals;
4) entrepreneurs, who use spatial data and plans to inform their business decisions;

5) residents, who use data about road and street closures or snowploughing to plan their trips effectively;
6) Members of Municipal Council use financial data in their decision-making processes.

It's worth noting that while some respondents mentioned the Members of the Municipal Council as users of the data shared on their website, they can be considered internal users rather than third parties.

Some of the interviewed organizations demonstrated awareness of examples where municipal open data was reused to bring public value. These instances included:

- applications in public transportation domain, such as bus schedules and real-time bus departure information;
- Waze that uses data on street and road closures to enhance their application's functionality;
- Minuomavalitsus.ee[10] that publishes a systematic overview of local governments' services, incorporating data collected by municipalities.

However, certain organizations lacked understanding of both reusers and use-cases, which emerged as a barrier, as discussed in later sections.

As previously mentioned, five of the interviewed local governments had published some data on the Estonian open data portal avaandmed.eesti.ee. At the time of writing (May 2024), one municipality from the sample had also published data on the national portal.

Datasets published by these municipalities on national portal range from common datasets including geospatial, budget data and document registers to specific sector information like transport, education, and planning, reflecting the varied needs and activities of different local governments. Geospatial data is shared by largest municipalities, namely Tallinn and Tartu, whereas smaller municipalities predominantly publish document registers. Notably, the municipality with the most datasets mentioned a recent reorganization of datasets on the national portal, combining those related to the same service. As such, the overall number seemingly decreased, however, the content remained unchanged. At the time of the interview, they had three types of datasets on the portal: (1) "classical machine-readable formats" as comes from respondents (e.g., JSON, CSV, XML), (2) spatial data for GIS software (e.g., map layers), and (3) historical data that is not updated (e.g., annual budget of a specific year as PDFs).

The main **driver for sharing data on the national portal** was often the initiative of key personnel within the municipality, and in most cases the first datasets were published already years ago. Two of the respondents admitted to not updating datasets since their initial disclosure and lacking awareness of the data-sharing process. I.e., LG1 mentioned that: *"...this became relevant for a moment when there was this open data portal, but no one has dealt with it more systematically… at least for about a year and a half this topic has not been relevant at all."*

**Challenges faced by local governments in sharing data** on the Estonian national portal included:

- no possibility to get datasets usage statistics (e.g., downloads in the last month/year). The overall statistics page of the portal is not usable – not possible to sort by data owner or dataset name;
- time-consuming metadata filling process – organization has to fill all metadata fields when adding a new dataset (no pre-filled fields);
- inability to convert data to various formats within the portal, while different user groups need different format of the data;
- lack of preview functionality for uploaded dataset;
- difficulty in identifying valuable datasets for publication.

---

[10] https://minuomavalitsus.ee/en

Seven municipalities had not shared any data on national portal, with four citing lack of **awareness or resources as the main reason**. For example, when elaborating on the possible reasons of lack of awareness of national portal existence, LG10 mentioned: *"There are so many topics here that perhaps some letters may have been overlooked or a notification has gone missing. I don't know of any other reason."* LG3 brought out: *"… we don't have the so-called resources or capability today to undertake this work, to gather these things together and start publishing them. In our house, it's more like this… the problems are on the objects, if there's a fire today, you deal with putting out that fire, rather than looking at how to possibly prevent future fires in this context - attention is paid to this, but rather less."*

There was an interesting case with one municipality that believed it was already sharing data on the national portal, as they had communicated with the Ministry of Economic Affairs and Communications, provided the necessary information, and assumed it would be published on the portal. However, no data had actually been shared. This situation highlights a misunderstanding of roles within the OGD ecosystem. Local governments often assume that data sharing is centrally managed by the Ministry, while the Ministry expects local governments, as the data owners, to handle the publication themselves.

Another reason mentioned by LG12 is that *"Many open data of municipality are today, for example, based on information systems and these information systems are located at service providers. And getting the data from there is also a bit problematic, because this is not foreseen in service level agreements."*

After general understanding of organizations experience with the OGD until now has been established, let us discuss specific barriers as implies form the model.

### 4.3 Usage barriers

Most of the usage barriers can be indicated as relevant for majority of the municipalities when openly sharing data.

Half of the respondents (6) agreed that the **inappropriate quality level of the municipality's data (UB1)** can be one of the barriers of openly sharing data. 2 respondents mentioned that the barrier lies more in the fact, that the quality level of the data is unknown or undefined. Two municipalities thought they may have quality issues in some operation fields and 2 of the respondents did not see quality as a barrier.

Majority of the interviewed municipalities (8) assessed the **process to prepare data for sharing as complicated (UB2).** 3 respondents, in turn, do not consider it complicated and one interviewee could not evaluate it due to lack of knowledge about this process. For the reasons for complexity of data preparation process, LG10 mentioned that in many cases they are doing manual work with no means of process automation, as well as lack of data management system. LG12 added that: *"…the information systems provide data differently and they have not taken it into account."* LG3 raised the competence issues: *"If there are no competencies and awareness in the organization, for example about the national portal, then it is difficult because someone has to start understanding it from scratch."* LG7 indicated the process itself is not complicated, but it is difficult to make data owners to implement and follow this process.

The **process to make municipality's data reusable by others (UB3)** was found to be complicated or somewhat complicated by 8 respondents, but the reasons varied quite a lot. In more detail, they were:

1) anonymization that in some cases requires manual or tailored solutions (LG10);
2) difficult to select the most appropriate format for data further reuse, e.g., JSON, CSV or XML (LG12);
3) time required to be dedicated to preparing a through documentation/description of the data, if done thoroughly (LG4);
4) limited knowledge and understanding of what is meant by "reusability" (LG3, LG5).

Two respondents assess the process itself not to be complicated, but "…*rather the concern is that we have no knowledge who is the user of this data*" (LG11) and "…*it is hard to find out what this necessary amount of data is that someone could actually use*" (LG4).

The next three barriers (UB4, UB5, UB6) concerned the open data portal - avaandmed.eesti.ee (i.e., all further responses to be treated as referring to national portal).

5 interviewees could not assess whether the **process to publish data on an open data portal is complicated (UB4)** as they had no knowledge or previous experience with it. One respondent mentioned that they started to create an account in the portal, and with some questions related to the process to which they had no immediate answers and therefore they stopped at that point. Some other challenges the municipalities (LG4, LG6, LG7) had faced were the same with those already mentioned in previous sub-section, namely: (1) publishing is too time-consuming (preparing metadata by manually filling respective form); (2) difficulty in creating APIs and getting data moving between the systems; (3) preparation of data in appropriate formats.

Only four respondents (LG8, LG10, LG11, LG5) agreed that the **process of publishing municipality's data on an open data portal is unclear (UB5)**. 3 municipalities mentioned that there are good guidelines available in the portal and the process is quite straightforward, "*…but maybe if you are using APIs, that might require more knowledge, but overall, most people could manage*" (LG6). One municipality that has not published data there mentioned: "*I read the guide, according to the guide, it seems like you have steps that you complete and it is supposed to work, but since I haven't tested it, I don't know if it actually works.*"

LG12 raised a different concern: "*I have understood that whatever is put up [in the portal] will be published… there seems to be no control.*" LG8 and LG9 mentioned again the awareness as one of the key factors, stressing that "*… when this knowledge [about publishing in OD portal] reaches that level, that it's actually maybe really some things that need to be done, then we would have a lot more data available through this portal.*"

Half of the respondents admitted that they cannot comment on the **functionality of open data portal (UB6)**, as they had not used the Estonian OD portal. For example, LG5 mentioned: "*...why I haven't used it, maybe because I haven't understood the benefits of it. Just collecting data doesn't provide any value. No one can value data until what you have collected reflects something back to us.*" Two municipalities find the national portal satisfactory with enough functionalities. However, two other municipalities criticized several aspects, such as (1) the limited functionality of the search tool; (2) the current portal's non-compliance with the Data Description Standard[11], which sets requirements for interoperable data descriptions; (3) the presence of many datasets that refer users to another system, requiring them to open another system to actually access the data.

Majority of the interviewed municipalities (9) acknowledged that **lack of possibility to semi-automate process to openly share its data (UB7)** is a barrier. "*If it were much easier to share data in open data portal, we would publish much more*" (LG5) and "*...if there were interfaces, it would make everything easier for us*" (LG6). As most of the Estonian local authorities use standard solutions and information systems, two respondents (LG1, LG12) suggested that the interface with the portal should be between the standard solution already, so that same datasets of different local governments would be published through API.

The last question regarding usage barriers concerning **process to maintain data once published (UB8)** was understood differently by respondents – some of them kept in mind the maintenance of all the data municipality is managing and others referred to the data published in open data portal. As for the latter, LG7 mentioned: "*At the moment the portal has no functionality to set up storage deadlines for datasets and you have to track somewhere else that*". On the other hand, LG6 did not see it as a problem: "*In some cases, it's good to keep certain things there. They're not going anywhere, if the need should arise.*"

---

[11] https://www.kratid.ee/en/juhised

Other usage barriers that form a barrier for municipalities to openly share its data are grouped in Table 4.

Table 4. Other usage barriers

| Usage barriers | Description or examples of the barriers |
|---|---|
| Lack of resources | No distinguished function or person who is responsible for open data and/or data management (LG1) |
| | Lack of financial and human resources (LG2, LG3) |
| | In some areas there is no data and creating it from scratch is very time-consuming (LG2) |
| Limited knowledge and awareness | The meaning of open data is unclear (LG12) |
| | Users of OD can make statistics, but not conclusions based on data (LG5) |
| | The awareness about the national portal is limited, it is not visible enough for users (LG9) |
| | Overall data competences and awareness of open data is low (LG1, LG10, LG3) |
| The value of organization's OD unclear | Not clear, who is the user of local level open data and therefore the value is unclear (LG1, LG3, LG10) |
| The "best" format of open data not defined | There are many machine-readable formats available, and it is unclear, which format would be most valuable for the users (LG4) |
| Unclear, what to publish | Limited knowledge of the users' needs – what datasets are wanted by users, i.e., high-value datasets (LG7) |
| | Difficult to prioritize, which data should be published (LG1) |
| Technical solutions | Lack of solutions, that would make the data processing more convenient (LG10) |

## 4.4 Value barriers

Next, barriers related to organizational beliefs and values were discussed, focusing on overall views on open government data and the organization's specific datasets. Some respondents found it challenging to distinguish between these perspectives, leading to answers for questions VB2 and VB3 being more focused on the organization's open data.

Most respondents (9) agreed or somewhat agreed that employees in their municipality believe that **openly sharing government data is often not valuable for the public (VB1).** For example, LG11 noted: "*…Often the purpose of it is not seen, as to why or for whom it is needed.*" Conversely, three municipalities, who provided examples of use-cases they were aware of in the first section of interview were more positive about it: *"...I actually think that the value of data to the public as a user is understood"* (LG9).

More than half of municipalities (7) acknowledged that people in their organization tend to believe that **many open government datasets are not appropriate for reuse (VB2).** Some reasons mentioned by respondents were that: (1) some data becomes useless when anonymized (LG10); (2) it is hard to see, who could be the user of a specific data organization owns (LG9); (3) there is a belief that nothing useful can be done with open data (LG3). 3 respondents were neutral about this question and 2 interviewees did not see this as a barrier, for example LG7: "*…considering that local governments operate at the level of power closest to the individual, on the contrary, opportunities may arise to value the data, which affects the daily lives of as many people as possible.*"

Nearly all municipalities (10) agreed that the organization believes that **many open government datasets suffer from data quality issues** (e.g., completeness, accuracy, uniqueness, consistency, etc.) **(VB3).** The most mentioned issues were inconsistency and inaccuracy. One example from LG8 states that "*The average salary or workload of teachers is a big mess…if the source data is incorrect, then it is not possible to get correct results.*" LG9 brought out that the reason for this belief might be the fact that people in organization tend to be too critical about their data. LG7 complemented this discussion with the following: "*...as far as I know, few agencies, organisations, companies have made quality determinations to their databases...so, if you do not know the quality, that means it is of questionable quality.*"

Most respondents (8) rather agreed (but not strongly) with the belief in organization that **public gains of openly sharing government data are often lower than the costs (VB4)**. For example, LG6: "*In terms of human resource cost… so much time is spent on it, but at the same time this information is not consumed, nothing is created based on it, so the question arises, why is it needed*" and LG10: "*…you are doing a lot*

*of work within that, but what do you do with that data after that… it is just nice, that you're getting some kind of overview, but well, what's next."* One municipality (LG4) argued that it is hard to have a straight answer, as some datasets are more valuable and interesting than others, but if the aim was to publish everything, then the costs would be too high compared to the public gains thereby pointing to the issues associated with data prioritization.

Only four municipalities believe that the **organization's gains of openly sharing government data are often lower than the costs (VB5).** For example, LG3: *"… if it were the other way around, if great benefits were seen, it would probably already be dealt with."* Three respondents had no opinion on this.

Half of municipalities agreed that **data preparation is too resource-consuming for their organization (VB6)**, as *"you have to process the data and it takes time"* (LG10). Two respondents (LG9, LG8) brough out that today it is not too resource-consuming, but they were not sure about sharing data on national portal, as they had no previous experience with it. One municipality mentioned that the biggest issue, i.e., the process that takes most resources is the data management, not sharing open data.

Half of the local governments rather agree with the belief that **open government data do not provide any value to their organization (VB7)**. For example, LG12: *"...people have not understood the value of open data"* and LG3: *"...the common understanding is that if you do something, direct return is important… indirect benefits don't tend to be seen"*. Some respondents commented that it could be the case in some fields or departments, but not a general belief in their organisation.

Most of the respondents (7) acknowledged that people in their local authority tend to believe that **open data that they can openly share will not provide value to users (VB8)**. Two municipalities mentioned that geoinformation may be an exception here, as the officials see more value in it. LG7 commented: *"...rather, there's no such belief, but very much would help if we knew more, what data users need."* LG 2 highlighted another aspect that officials in the organization that do not work with open data cannot see its value.

The final value barrier - the **number of resources to be spent to prepare, publish and maintain open government data outweigh the benefit my organization gains from it (VB9)** – lead to ambiguous responses. Only 4 municipalities totally agreed with it. Three respondents rather disagreed and five of them did not have a clear answer. For example, LG10: *"To a certain extent, I think benefits are seen, but if it starts to take most of the working time, it will go the other way."* LG9 points out that: *"I rather think that this is not the case…we overestimate the effort we must put into it and underestimate the benefit. The benefit of such things is often underestimated, as it is often indirect."*

The other value barriers that respondents see as a barrier for their municipality to openly share its data that were not covered by the model are provided in Table 5.

Table 5. Other value barriers

| Value barriers | Description or example of the barrier |
| --- | --- |
| Duplicated data collection by different agencies | Same kind of data is gathered by different agencies – brings the question, which of them is reliable or which data to use as a basis. E.g., in case of residents of municipalities, the data of Estonian Population Register and Statistics Estonia is different (LG10) |
| Value of open data portal unclear | The value of Estonian open data portal is unknown to public (LG12) |
| | Anonymized data in some cases is useless and can only be used for statistical purposes (LG2) |
| Terms "data" and "open data" are unclear | Understanding of the term "data" and "open data" is different, not uniform and needs clarification amongst the officials (LG3, LG7) |
| Open data not prioritized | Open data is not a priority for organization (LG8, LG10) |

## 4.5 Risk barriers

Risk barriers pertain to the uncertainty regarding the financial, functional, and social consequences of using or publishing open data. Most of the risk barriers of the model were found relevant or somewhat relevant by the respondents, with only one barrier deemed not applicable.

Half of the respondents (6) agreed that the **fear of the misuse of openly shared government data (RB1)** may be a barrier. Three municipalities raised concerns about security, not only criminal activity, but also cybersecurity, which is a new topic in regards the situation in the world. On the contrary, some respondents argue that open data *per se* means that the possibilities of misuse are already minimized.

The **fear of misinterpretation of openly shared government data (RB2)** was somewhat more often acknowledged, as 8 respondents raised different concerns connected to it. For example, LG2: *"Many officials also fear that if we publish too much, people will start asking too much or get too aware… I don't see fear in it, but, well, some do"*. Some gave the example of financial data and documents, that might be the source of misinterpretation as people have limited understanding of this data. Conversely, two respondents mentioned that if someone makes a false interpretation or draws an incorrect conclusion from the data, it reflects the individual's limited understanding rather than an issue with the data itself. LG7 commented: *"…this possibility exists, but we try to minimize it."*

6 municipalities agreed that this might be the case in their organization that people **fear that openly shared government data will not be reused (RB3)**. For example, LG3 pointed out that: *"…it could be from this view that who needs them at all and why we should do it."* Four respondents did not consider this fear a barrier to openly sharing data. For example, LG6 mentioned: *"…I think we are not afraid of this, as we have offered data and it is up to users to do something with it"*. Two municipalities mentioned that the "fear" is not relevant, but "belief" that OGD will not be reused is relevant (covered in previous section on value barriers).

The majority of interviewed municipalities (10) acknowledged the **fear of violating data protection legislation when openly sharing government data (RB4)**, particularly among officials handling personal data. Respondents highlighted several reasons and additional concerns: (1) the boundaries of what constitutes personal data are delicate or blurred (LG11, LG4); (2) fear of injunctions by the inspectorate (LG12); (3) hesitancy to process data due to data protection concerns (LG4); (4) previous cases or negative experiences have made people cautious (LG7); (5) document management systems are primary sources where data that should not be openly shared could be exposed (LG7, LG9). Two municipalities expressed confidence that this fear is not widespread in their organization, as there is a clear understanding of what data to restrict.

The respondents were more ambivalent about the **fear that sensitive data will be exposed as a result of opening its data (RB5)**. Half of them agreed with it and some referred that it is connected to the previous question, and this might be relevant to their organisation. But some respondents commented that sensitive information is usually removed from data during the preparation of opening data and it cannot be the case that, for example *"…data of disabled persons is accidentally exposed"* (LG4).

Less than half of the municipalities (5) agreed or somewhat agreed with the **fear of making mistakes when preparing data for publication (RB6).** For example, LG10 mentioned: *"…it is also feared, but maybe less than compared to the previous two [fears]"* and LG1 complemented: *"…as we cannot be sure about the quality of our data, then the fear is automatically there."* LG7 pointed out, that this is not a widespread fear, but some people are more worrying. Two respondents did not see it applicable to their organization, as this fear could be relevant if they would process and share more data than today.

The **fear that users will find existing errors in the data (RB7)** received mixed feedback. One respondent mentioned that it is probably the biggest fear related to data (LG5) and two municipalities acknowledged that this fear might be relevant, but not to the extent that it forms a barrier to share data (LG1, LG4). This fear could be related to the problems with quality of data (LG11) and human fear of making mistakes (LG3, LG6). On the contrary, nearly half of respondents held the opposite perspective, for example LG12: *"…if somebody finds mistakes, it's good, we'll find out there's something wrong. That's the feedback, that's good,"* and LG8: *"…it's overall positive, then we know that somebody uses it."*

Majority of municipalities do not find the **fear that openly sharing its data will reduce its gains (RB8)** relevant as they cannot see the possibility, how municipality could sell the data or whether it is even possible by law. One respondent provided an interesting example, how companies are gathering data of procurements from municipalities' websites and then offer this data for the service provides, who have joined their procurement/tender platforms.

The **fear that openly sharing organization's data will allow its competitors to benefit from this data (RB9)** was found irrelevant, as municipalities are public sector bodies and could not see, who the competitors in their territory might be. The local governments are watching each other's initiatives and, in some cases, compete over residents, but in terms of data there is no competition.

Other risks or fears mentioned by respondents that might form a barrier for the municipality to openly share its data are provided in Table 6.

Table 6. Other risk barriers

| Risk barrier | Description or example of the barrier |
| --- | --- |
| Security issues | Publishing of some type of data might be used for criminal motives (e.g., cyber-attacks, security breach) (LG1, LG2) |
| General fear of the unknown | LG3: "Lack of data awareness creates fears, as everything unknown seems scary to some people" |
| Lack of resources | Overall lack of resources (LG7) |

## 4.6 Tradition barriers

Tradition barriers were considered somewhat less relevant by respondents compared to usage and value barriers. The interviewed municipalities split in thirds in responses regarding the belief that **Freedom of information requests are sufficient for the public to obtain government data (TB1)**. The respondents who were subtle commented, that *"...this might be the case of some fields of activity, but not a widespread belief"* (LG5). Two municipalities brought out that if enough data is already published, then there are less information requests, and it means less work with them.

The **reluctancy to implement the culture change required for openly sharing data (TB2)** was not seen as a barrier by most of the respondents. LG9 and LG10 pointed out that major cultural shift was done during the administrative reform[12] and LG3 commented they are on the transition point, where the readiness for change is somewhat there, but the traditions also prevail. LG5 commented: *"It depends greatly on management and politicians...in our case there is rather desire to publish more"*. LG2 and LG7 admitted that there might be some people in organization who are reluctant to change, but this cannot be applied to the entire organization.

Half of the respondents (6) agreed that **employees in their municipality lack the skills required for openly sharing data (TB3)**. Three municipalities thought that the basic skills exist, but there might be some specific skills that need development. LG1 broadened the issue: *"...starting with the notion of what value the data carries... it is viewed as only in the context of their own department and no one can assess in the context of an organization or the customer point of view."* Some respondents referred to the national OD portal and admitted that they have no experience with it, but it would probably be only the first time a new thing to learn.

The respondents interpreted the question regarding the **lack of the skills required for maintaining openly shared data (TB4)** differently. Most referred to the overall data maintenance and did not perceive it as a problem, as it aligns with their daily practices. However, some municipalities, particularly those responding from the perspective of maintaining data in the national portal, acknowledged that manual maintenance requires someone to oversee the data (e.g., ensuring it is up-to-date), which consumes time.

Most municipalities rather would not agree that **organization is reluctant to radically change the organizational processes that would enable openly sharing government data (TB5).** The arguments were following: (a) the management is positive about the change, but some employees are reluctant (LG10), (b) the desire for change is there, but lack of financial and human resources are the barriers for change (LG3, LG11), (c) open data is not valued and not prioritized (LG5).

---

[12] Estonian administrative reform was implemented in 2015-2017, for further information access: https://www.agri.ee/en/administrative-reform-2015-2017

Other tradition barriers mentioned by respondents, which might hinder the municipality from openly sharing its data, are detailed in Table 7. Interestingly, while most municipalities did not acknowledge reluctance to change in culture or processes as a barrier (TB2), several of them emphasized during the discussion of other tradition barriers the general resistance to changes within the organization. This included negative attitudes towards any kind of changes among the personnel, indicating a broader cultural reluctance to embrace new approaches or initiatives.

Table 7. Other tradition barriers

| Tradition barrier | Description or example of the barrier |
|---|---|
| Open data is not a priority | There are main functions that people are used to doing, open data is something new (LG1) <br> Open data is not valued (LG5) |
| Reluctance/resistance to change | Habit of doing things the same way as done previously (LG2) <br> Negative attitude towards change (LG3) <br> It is difficult to change the mindset of people, who are used to do things as always (LG5) |
| Technophobia - irrational fear of technology | Fear that digitalisation brings more work than savings (LG9) |
| Lack of data management processes | Estonian data management Framework is difficult to understand and implement at local level (LG6) |

## 4.7 Image barriers

Majority of the interviewed municipalities (8) consider that their **organization do not have negative image of open government data (IB1)**. Two respondents mentioned that the image is neutral, as LG10 commented: *"[The negative image] may be a case for some data…as time goes on, the more people want their work to have a meaning…time resource versus the benefits of it is being considered in their work."*

Half of the respondents agreed or rather agreed that their organization **believes that open government data is not valuable for users (IB2)**. 4 municipalities rather did not agree, they acknowledged that there might be a few people who believe it (LG6) or it might be relevant in case of some datasets (LG4). LG8 commented: *"…rather they fear it or do not know, whether the data is valuable or not."*

The municipalities split into two regarding the **fear that openly sharing government data will damage the reputation of their municipality (IB3)**. Half of them disagreed, for example, LG7 commented: *"…rather, we strive for such a reputation that it shows us in a better light."* The other half of respondents rather agreed that: (a) this fear might be the case for some data (LG1), (b) it could be connected to the fear of making mistakes and data quality issues (LG10), (c) data can reveal the "real picture", which is not as good as expected (LG3, LG9). LG5 pointed out that some employees fear that open data related to their work might show them in a poor light.

Majority of the respondents (9) admitted that the **fear that the accidental publication of low-quality data will damage the reputation of their municipality (IB4)** may be applicable to them. Two of them specified that this is more a risk they must mitigate.

Half of the municipalities acknowledged that their organization **fears that being associated with incorrect conclusions drawn from OGD analysis by OGD users will damage its reputation (IB5)**. Examples provided by respondents include instances of (1) wrong conclusions based on document registers (LG10), (2) media making incorrect interpretations (LG11, LG9), and (3) the challenge of disproving such incorrect conclusions (LG9).

Most interviewees did not see any other image barriers for their municipalities to openly share its data. But LG9 raised an interesting point: *"The topic of open data is not one that receives significant public credit, and therefore, it is not given excessive attention. …if it were something exciting, there might be a desire to support those who are considered cool… this kind of order would probably change a lot and a lot of data would come on the market."*

## 4.8 Other barriers

Half of the respondents identified additional barriers beyond those covered by the model utilized for the interviews, primarily falling into the categories of usage and value barriers. However, some highlighted barriers are not initially included in the model, which could inform its refinement:

1) lack of data awareness – individuals working in local authorities may lack awareness of the concept of open data or have an overall understanding gap regarding data;
2) lack of data competence – there may be insufficient knowledge and skills to work with data, such as data governance or management, data analytics, and data quality management;
3) lack on human resources - there may be no designated function or person responsible for open data and/or data management;
4) lack of financial resources – there might be an overall shortage of funds to fulfil all obligations imposed on local governments;
5) lack of clarity on what data to publish – uncertainty exists regarding which data would be most valuable for users;
6) difficult in prioritizing data - challenges arise in determining which data should be published first.

## 4.9 Open data ecosystem

The final part of the interview covered overall open data ecosystem – its actors, the benefit of open data, measures to improve the current situation and the awareness of policy documents and initiatives.

All the respondents were able to identify at least two **open data ecosystems actors**: the data owners (e.g., municipalities, state agencies) and the data users/ consumers (e.g., private sector, academia, journalists, other public agencies). Additionally, half of the interviewees mentioned some form of mediator or intermediary – such as a platform, portal, or solution – that facilitates consumer access to, viewing, downloading, or requesting data. Two municipalities identified another key actor - regulatory body and Data Protection Inspectorate, responsible for creating regulations and ensuring compliance. Furthermore, one respondent highlighted stakeholders who stand to benefit from open data initiatives.

The majority of top-level management respondents indicated a lack of awareness regarding **policy documents or initiatives guiding open data development** at either the Estonian or European level (e.g., Estonian Digital State 2030). However, some of them specified that employees responsible for this domain likely possess knowledge about them. Additionally, three municipalities pointed out that these documents often inadequately address the needs of local government levels, portraying them as more akin to slogans or aiming to enhance the digital state's reputation.

Several recommendations were made regarding **actions to be taken at the national level** (e.g., Ministry) to facilitate municipalities in disclosing more open data and adhering to related requirements:

- **defining most important or valuable datasets that municipalities should prepare and share** on the national portal (LG1, LG12, LG6) serving as a pivotal starting point for open data initiatives at the local level;
- resolving the issues concerning the revenue base of local governments , as it is challenging to contemplate necessary changes when basic municipal services lack resources (LG11, LG3);
- resolving problems with compatibility and interfaces must such as data duplication across different databases and manual insertion, with the need to streamline data management processes (LG10);
- more practical guidance, training and assistance tailored to municipalities from the state agencies, rather than solely offering general documentation and informational events (LG12, LG11, LG6, LG7);

- implementing direct application programming interfaces (API) between standard software used by local governments and the national open data portal to enhance data exchange efficiency (LG12, LG5);
- showcasing the benefits and value of municipal open data (LG5, LG1);
- defining specific measures for municipalities that would incentivize them to publish more open data defined (e.g., the portal minuomavalitsus.ee[13]).

The majority of respondents (8) agreed that average and smaller municipalities are likely unable to handle this topic on their own and emphasized the inevitability of cooperation among them. They suggested several **actions that could be taken by local governments**:

- implementing uniform information systems across municipalities to ensure consistent data collection and management practices (LG5, LG12);
- enhancing awareness among local government employees about the importance and benefits of open data initiatives (LG2, LG9);
- facilitating knowledge exchange platforms for sharing insights and advancements in data analytics developments among municipalities (LG1);
- ensuring the proper implementation of data management and governance frameworks to streamline the process of publishing open data (LG5, LG6, LG7);
- providing guidance or training on open data initiatives through associations such as the Association of Estonian Cities and Municipalities (LG8, LG11, LG12);
- harmonizing or standardizing datasets related to similar municipal functions to ensure interoperability and facilitate cross-municipality data usage (LG4);
  integrating data management practices with strategic management processes to enhance the use of data in decision-making within municipalities (LG5).

Based on the barriers identified in the previous subsections, the first priority should be to raise awareness and build competence regarding municipal data. As highlighted by LG2: "*Awareness needs to be significantly increased to truly understand the data that local governments produce themselves. That's where the focus should be directed…it doesn't make sense to hold or duplicate data that already exists elsewhere in the country, which I see happening frequently. By doing so, the work itself becomes more focused*".

Despite the challenges faced by municipalities and the barriers they experience within the studied context, and often a lack of awareness of the value of data, many municipalities do recognize the **benefit or value of open data**, which are provided in Table 8.

Table 8. Opinions on benefit or value of open data

| Benefit or value | Description or example |
| --- | --- |
| Increase of service quality | The quality of some services might increase due to the analysis made by combining open data (LG1) |
| | Use of geospatial information by architects, planners etc. makes processes faster and of higher quality (LG2) |
| Public value creation | Even if the value is not seen now, there might be someone there who already foresees it (LG7) |
| New services | New services created by private or third sector (LG4, LG6, LG7) |
| Proactive services and prevention activities | Linked data could be a basis for proactive services and prevention activities, e.g., road closures in Waze application (LG10) |
| Smarter decisions | Open data can be helpful in the decision processes, e.g., designing public services or business (LG8, LG12) |
| | The biggest value comes from knowledge and correct/accurate conclusions that are drawn on data (LG3, LG5, LG9) |
| Less referrals from residents | Residents have more data available, and the number of referrals decreases because of it (LG2) |

---

[13] https://minuomavalitsus.ee/en

| Prevent the spread of misinformation | It is possible to rely on data in case of intentional dissemination of misinformation (LG3) |
| Rise of data awareness | Enables to increase data awareness and data literacy, e.g., data analysis skills (e.g., already on lower levels of education) (LG4) |

## 4.10 Recommendations for ecosystem improvements

Based on interviews with local governments, several barriers considered most relevant by respondents were identified. From these barriers, we derive recommendations to overcome or mitigate them, as suggested by interviewees.

The results of the interviews revealed 18 barriers of the IRT model with which at least half of the respondents agreed. The highest number (6) was related to value barriers, 4 to both risk and usage barriers, 3 to image barriers, and 1 to tradition barriers. Consequently, all categories of IRT barriers are relevant for local governments, albeit to different extents. Table 9 presents (1.1) the barriers not included in the initial model as identified from interviews and (1.2) the most relevant barriers already included in the model, where barriers addressing similar or overlapping issues are combined. Additionally, it provides (2) the recommended measures for ecosystem improvement at (2.1) national level (e.g., policymakers, portal owner) and (2.2) local level (individual or local governments collectively).

Table 9. Most relevant barriers and recommendations for improvements

| Barrier | Measures at national level (e.g., policymakers, portal owner) | Measures at local level (individual LG or collectively) |
|---|---|---|
| Lack of data awareness/ competence<br>Lack of skills for openly sharing data (TB3) | Practical guidance, training and help for municipalities | Guidance and training offered by Association of Estonian Cities and Municipalities, knowledge exchange about data governance and analytics developments, e.g., best practices |
| Lack of human and financial resources | Revision of the revenue base of local governments<br>Specific measures for municipalities that would motivate them to allocate resources there | Increase awareness of existing employees, identify "data enthusiasts" within organization |
| Lack of clarity, what data to publish | Centrally agree upon the most relevant/valuable datasets, that local governments must publish themselves (e.g., initiated by ministry in cooperation with local governments) | |
| Inappropriate data quality (UB1, IB4, VB3) | Adaption of central data governance guidelines (including data quality) for municipalities | Implementation of data management/governance & automated data quality management procedures |
| Complexity of data preparation (for reuse) (UB2, UB3)<br>Resource intensity of data preparation (VB6) | Solutions that would make the data processing more convenient (augmentation and automation), enhancement of OD portal | Publishing of data in formats that require minimal additional work |
| No semi-automate process to publish data (UB7) | Direct application programming interfaces (API) between standard software (used by local governments) and national open data portal | |
| Lack of OGD value awareness for public/ users (VB1, IB2, VB8) | Benefits and value of municipal open data should be evidenced by use-cases; make usage statistics of datasets available to publishers | Uniformed information systems that provide more uniformed data |
| Datasets not appropriate for reuse (VB2) | Datasets related to the same municipal functions could be harmonized/uniformed by defining uniform granularity level | |
| Public gains lower than costs (VB4) | Demonstrate the impact of open data through showcases/ use-cases/re-uses, success stories | Testimonials about OGD use-cases, |
| Fear of misuse and misinterpretation (RB1, RB2) | n/a | n/a |
| Fear of violating data protection regulations and exposure of sensitive data (RB4, RB5) | Clear guidelines and examples of different cases | Implement processes that mitigate the risks |
| Incorrect conclusions drawn from OGD (IB5) | Improvement of overall data analytics skills (e.g., in general education, adult training) | |

For certain barriers (designated as n/a in Table 9), specific countermeasures or activities were not identified during the interviews with local governments. This particularly applies to fears of misuse and misinterpretation (RB1, RB2), for which potential measures to alleviate concerns might include: (1) enhancing overall data literacy for all actors in the open data ecosystem through initiatives such as seminars, workshops, and courses, (2) establishing feedback mechanisms and platforms for the exchange of best practices, and (3) developing and promoting awareness of robust data governance policies.

Moreover, the qualitative analysis uncovered the primary user groups and some use cases of open data at the local level known to municipalities, including students, journalists, entrepreneurs, residents, spatial planning actors, and council members. This insight could serve as a valuable foundation for fostering collaboration among municipalities and conducting a more comprehensive analysis of user needs.

## 5 Discussion and Limitations

While Estonia has demonstrated progress in open data rankings, there are still notable areas for improvement, particularly at the local government level. Despite major cities like Tallinn and Tartu are publishing their data as OGD, the majority of municipalities have yet to contribute to the national portal. This discrepancy underscores a broader issue of limited awareness and understanding of open data among both the public and local administrative bodies. Despite the recognized importance of open data, there is a noticeable lack of research focusing on OGD at the local government level, not only in Estonia but also internationally. Acknowledging the growing significance of local and smart city open data portals within the broader ecosystem (Liva et al., 2023; Lnenicka et al., 2024) and a noticeable lack of research focusing on OGD at the local government level (Fusi, 2021; Sandoval-Almazan et al., 2021), not only in Estonia but also internationally (Rajamäe-Soosaar and Nikiforova, 2024), it is imperative to address the current limitations in local open data provision and to enhance the visibility of local OGD in the national portal.

To address this gap, this study aimed to identify the barriers preventing Estonian municipalities from openly sharing data and to propose effective strategies to overcome these obstacles. To this end, qualitative analysis was conducted, employing the OGD-adapted Innovation Resistance Theory model, through interviews with 12 local governments. Results highlighted that while municipalities collect diverse data types, various challenges in openly sharing data on persist.

Challenges faced by municipalities include human resource constraints, technical platform constraints, lack of data competencies, and difficulties in preparing data to be shared as open data due to the format in which they are stored. Technical platform constraints associated with the Estonian open data portal (1) the inability to get usage statistics for datasets, (2) the time-consuming process of filling metadata fields with lack of automation of the process, (3) difficulty in creating APIs and getting data transferred between the systems, (4) preparation of data in appropriate formats, (5) difficulty in prioritizing and identifying high-value datasets for their further publication. More than half of the municipalities interviewed had not yet shared any data on the national portal at the time of the interview, primarily due to either a lack of awareness or insufficient resources. However, following the interviews, some municipalities indicated their intention to address this issue soon or to allocate more attention to it within their organizations. Another challenge highlighted was the difficulty in accessing data from information systems hosted by service providers, which was not foreseen in service level agreements.

The barrier types covered by the IRT model include various usage, value, risk, tradition, and image barriers. The most relevant barriers for openly sharing the data include: (1) lack of awareness and skills for open data sharing (TB3), (2) insufficient human and financial resources, (3) uncertainty about what data to publish, (4) concerns about data quality (UB1, IB4, VB3), (5) the complexity and resource intensity of data preparation for first-time publishing and further re-use (VB6, UB2, UB3), (6) lack of data publishing process automation (UB7), (7) lack of understanding of the value of open data for the public/users (VB1, VB2, VB8, IB2) and belief that public gains are lower than costs of data opening (VB4), (8) fears about misuse and misinterpretation of opened data (RB1, RB2), (9) fear of potential violations of data protection regulations (RB4, RB5), and (10) the possibility of incorrect conclusions being drawn from the data (IB5).

The interviewees proposed several measures to address these barriers at both national and local levels. These recommendations could serve as input for developing specific action plans to assist local governments in navigating the challenges of open data. For instance, interviewees suggested conducting awareness-raising activities and providing practical guidance and training, strategies that have also been supported by previous studies (Golub and Lund, 2021; Kawashita et al., 2022). Additionally, they highlighted the importance of automation and implementing motivation measures for local governments.

Furthermore, the analysis uncovered that while data governance guidelines have been established by the central government, they often lack applicability to municipalities and necessitate adaptation to local contexts through practical sample procedures. Solutions that streamline data processing, such as augmentation, automation, and APIs, were also recommended to reduce manual workload.

Given that the value and benefit of open data emerged as the most significant barrier, there is a pressing need to emphasize its importance by showcasing evidence through use-cases, testimonials, and success stories. This aligns with the findings of previous research (Lassinantti et al., 2019; Lnenicka et al., 2024; Ansari et al., 2022; Hein et al., 2023; Chokki et al., 2022), which underscores the importance of understanding the various types of data users and their motivations in leveraging open data for valuable impact. Moreover, making detailed usage statistics and analytics of datasets on the national portal available to data publishers would enable them to monitor user interest effectively.

One of the key measures to address the current situation involves reaching a consensus on the most relevant and valuable datasets that local governments should publish, a notion supported by existing literature (Kević et al., 2023; Utamachant & Anutariya, 2018; Nikiforova, 2021). Additionally, it is crucial to address the security aspect of open data, particularly concerning spatial data (in line with (Yang et al., 2015). These actions necessitate collaboration between local governments and the national level, including the ministry. Harmonizing datasets related to specific services by defining uniform granularity levels and data dictionaries could further streamline the process and contribute to the interoperability of published data (in line with (Hein et al., 2023; Kawashita et al., 2024; Ali et al., 224)).

The prevalent fear among municipalities regarding potential violations of data protection regulations (in line with (Li et al., 2024)) underscores the importance of clear guidelines and the implementation of processes to mitigate associated risks. Active monitoring by the Inspectorate and adherence to GDPR regulations are essential steps in this regard.

However, this study has several limitations. Firstly, the primary findings rely on qualitative analysis derived from interviews with local stakeholders, inherently introducing subjectivity into the results. However, this methodological choice was deliberate, as qualitative approaches offer a deeper understanding of the subjective experiences and contextual nuances that quantitative methods may overlook and help to identify unforeseen factors that may influence the phenomenon under investigation.

Secondly, the sample size is constrained to 15% of Estonian municipalities (12 out of 79), potentially impacting the generalizability of the findings. Future research endeavors could benefit from employing a quantitative approach that encompasses all Estonian municipalities, thus offering broader insights and enhancing the objectivity of the conclusions.

Thirdly, since the research focuses specifically on Estonian local governments, the results may not be directly applicable to other contexts with differing administrative systems. It is essential to interpret the findings within the context of Estonia's unique governance structure and consider potential variations in other countries.

Lastly, given the dynamic nature of the topic, the results reflect a snapshot in time, emphasizing the immediate relevance of their application in enhancing the local data ecosystem and associated data provision. Furthermore, it is recommended to replicate the study in the future to identify any remaining barriers or new challenges that may have emerged. By using the presented results as a baseline, future iterations of the study can evaluate the impact of interventions and assess changes in perceived barriers over time.

## 6 Conclusions and future research directions

This study aimed to examine the barriers hindering open data sharing within local authorities in Estonia. By focusing on the specific context of local government data sharing practices, we contribute to the broader understanding of open data governance and implementation, particularly within the context of local data ecosystem, which is an integral part of the open data ecosystem.

The core contribution of this study lies in its examination of the challenges faced by Estonian local governments in embracing open data initiatives. By delving into the reasons behind the disparity between central government and local authority efforts, as evidenced by the Estonian Digital Agenda 2030, key impediments and actionable solutions were determined.

Through interviews with representatives from local governments, this research sheds light on the complexities surrounding data sharing practices at the municipal level. It not only uncovers the primary barriers inhibiting open data sharing but also highlights potential enablers that could facilitate smoother implementation. From enhancing awareness and skills among local government stakeholders to improving data governance frameworks and infrastructure, it offers a roadmap for creating a more conducive environment for open data sharing leading to more resilient and sustainable open data ecosystem. The study also revealed and communicated with interviewees some simple steps for improvement, e.g., publishing open data of local government document management systems on national portal with one of the municipalities published data after the interview.

Given that several databases the municipalities are using for data collection, are centralized or national, it is crucial to study, which municipal data should be published by local governments themselves and in which cases the central register/body should be the publisher. This requires cooperation between local governments and the national level and could be one of the future research directions. In addition, as the quality of open data and metadata were often named during interviews as potential barriers preventing data opening or complicating it, as well as data maintenance, the future studies could take this into focus, including considerations of integrating of respective mechanisms for their control and improvement into the OGD portal. In addition, the study implicated gaps in knowledge about the actual users and their needs, thus mapping the main user groups of local-level OGD and developing detailed use-cases of data reuse would be valuable to bring more comprehensive view. The qualitative analysis on local level OGD provision barriers and enablers could be complemented with quantitative analysis, to validate the relevant barriers identified by this study.

By validating the OGD-adapted IRT model, which has not yet been tested in real-world scenarios, and proposing refinements based on empirical evidence, this study contributes to the advancement of theoretical frameworks for understanding resistance to data sharing. Moreover, it has been demonstrated that the model is applicable at both the national level that it was originally developed for and local data ecosystem to which it was tailored in this study. As such, this study also proposed a revised version of the instrument tailored for local government context. As part of the model validation by its application to Estonian case, all barriers were found to be relevant, while in addition, municipalities raised some barriers not included in the initial model falling predominantly to the categories of usage and value barriers. These implications can serve as possible additions for the OGD-adapted IRT model refinement upon they are validated.

## Declaration of Generative AI and AI-assisted technologies in the writing process

During the preparation of this work the authors used Copilot in order to improve formatting, styling and translation. After using this tool/service, the authors reviewed and edited the content as needed and takes full responsibility for the content of the publication.

# Appendix

## Appendix A. Example of transforming usage barrier and associated items into interview questions

| Barrier and measurement items | Interview questions |
|---|---|
| **Usage barrier (UB)** | **Q12.** Are there any usage barriers related to the required changes in your municipality's routines that form a challenge for openly sharing your municipality's data? (UB) |
|  | **Q13.** To what extent do the following situations form a barrier to openly sharing your municipality's data: |
| **UB1:** It is difficult to attain the appropriate quality level for open government data to be shared openly | − an inappropriate quality level of your organization's data? **(UB1)** |
| **UB2:** It is difficult to prepare data for publication so that they comply with OGD principles | − a complicated process to prepare data for sharing? **(UB2)** |
| **UB3:** It is difficult to prepare data for publication so that they become appropriate for reuse | − a complicated process to make your municipality's data reusable by others? **(UB3)** |
| **UB4:** Data are difficult to publish on the OGD portal due to the complexity of the process | − a complicated process to publish your municipality's data on an open data portal? **(UB4)** |
| **UB5:** Data are difficult to publish on the OGD portal due to the unclear process | − an unclear process of publishing your municipality's data on an open data portal **(UB5)** |
| **UB6:** Data are difficult to publish on the OGD portal due to their limited functionality | − limited functionality of open data portals? **(UB6)** |
| **UB7:** Open government data portals often do not allow for semi-automation of the publishing process | − no possibility to semi-automate my organization's process to openly share its data? **(UB7)** |
| **UB8:** It is difficult to maintain openly shared government data | − the need and a complicated process to maintain data once published **(UB8)** |
|  | Q14. Are there any other usage barriers that form a barrier for your municipality to openly share its data? **(UBn)** |
| Value Barrier (**VB**) | … |
| Risk Barrier (**RB**) | … |
| Tradition Barrier (**TB**) | … |
| Image Barrier (**IB**) | … |

## Appendix B. Codes used in NVivo to analyse the interviews

| Name | Description | Interviews | References |
|---|---|---|---|
| Image barriers |  | 12 | 56 |
| IB1. Negative image of OGD | your municipality has a negative image of open government data | 10 | 11 |
| IB2. Belief that OGD is not valuable for users | your municipality believes that open government data is not valuable for users | 11 | 11 |
| IB3. OGD damages reputation of LG | your municipality fears that openly sharing government data will damage the reputation of your municipality | 10 | 10 |

| Name | Description | Interviews | References |
|---|---|---|---|
| IB4. Accidental publication of low-quality data damage reputation | your municipality fears that the accidental publication of low-quality data will damage the reputation of your municipality | 11 | 12 |
| IB5. Incorrect conclusions drawn from OGD | your municipality fears that associating them with incorrect conclusions drawn from OGD analysis by OGD users will damage its reputation | 9 | 9 |
| Other image barriers | | 3 | 3 |
| OD ecosystem | questions about overall open data ecosystem and potential improvements | 12 | 66 |
| Actions by LG individually or in cooperation | Which actions should be taken by local governments themselves to improve the publication of open data? (individually or in cooperation) | 12 | 13 |
| Actions on national level | Which actions should be taken at the national level (e.g., Ministry) to help municipalities disclose more open data and comply with the requirements related to open data? | 11 | 13 |
| Benefit or value of OD | What could be the benefit or value of open data in your opinion? | 11 | 20 |
| OGD ecosystem actors | How do you see the whole open data ecosystem – which actors and what does it include? | 10 | 12 |
| Policy documents or initatives of OGD | Are you aware of any policy documents or initiatives guiding developments in the field of open data on Estonian or European level? Are you aware of Estonian Digital State Development Plan? | 8 | 8 |
| Organization and municipal portal | | 12 | 75 |
| Cases of reuse of OD | specific cases, where your municipality's open data was reused by third party | 11 | 21 |
| Challenges of sharing data | What challenges did your municipality face in openly sharing data? | 11 | 15 |
| Drivers to share data on municipal portal | | 8 | 8 |
| Process of sharing data | how the process of openly sharing data worked and who within your municipality was or were involved? | 11 | 12 |
| Type of data on municipal portal | What type of data did your municipality share openly and how often? | 12 | 19 |
| Organization and national portal | | 12 | 28 |
| Cases of reuse of OD | specific cases, where your municipality's open data was reused by third party | 1 | 1 |
| Challenges of sharing data | What challenges did your municipality face in openly sharing data | 4 | 6 |
| Drivers to share data on national portal | | 3 | 3 |
| Process of sharing data | how the process of openly sharing data worked and who within your municipality was or were involved | 3 | 3 |
| Reasons for not sharing data | | 7 | 11 |
| Type of data on national portal | What type of data did your municipality share openly and how often | 3 | 4 |
| Organization of the interviewee | some questions about the organization you are working for | 12 | 33 |
| Role of respondent in LG | What is your role in this municipality | 12 | 13 |
| Type of data LG collects | What type of data does your municipality collect | 12 | 20 |
| Overall barriers | Are there any other barriers that form a barrier for your municipality to openly share its data (not limited to the categories above) | 9 | 12 |
| Overall understanding of OD | | 10 | 12 |
| Risk barriers | | 12 | 94 |

| Name | Description | Inter-views | References |
|---|---|---|---|
| Other risk barriers | | 8 | 8 |
| RB1. Fear of misuse of OGD | -the fear of the misuse of openly shared government data | 11 | 11 |
| RB2. Fear of misinterpretation of OGD | -the fear the misinterpretation of openly shared government data | 11 | 12 |
| RB3. Fear that OGD not reused | the fear that openly shared government data will not be re-used | 11 | 11 |
| RB4. Fear of violating data protection | -the fear violating data protection legislation when openly sharing government data | 12 | 13 |
| RB5. Fear that sensitive data is exposed | -the fear that sensitive data will be exposed as a result of opening its data | 9 | 9 |
| RB6. Fear of making mistakes | the fear making mistakes when preparing data for publication | 9 | 9 |
| RB7. Fear that users find errors in data | the fear that users will find existing errors in the data | 11 | 11 |
| RB8. Fear that OGD reduces gains | the fear that openly sharing its data will reduce its gains (otherwise the municipality could sell the data or use it in another beneficial way) | 6 | 7 |
| RB9. Fear that OGD allows competitors benefit | the fear that openly sharing its data will allow its competitors to benefit from this data | 3 | 3 |
| Tradition barriers | | 12 | 65 |
| Other tradition barriers | | 7 | 9 |
| TB1. Information requests are sufficient | your municipality believes that Freedom of information requests are sufficient for the public to obtain government data | 11 | 11 |
| TB2. Reluctancy to implement culture change | your municipality is reluctant to implement the culture change required for openly sharing government data | 12 | 13 |
| TB3. Lack of skills for openly sharing data | employees in your municipality lack the skills required for openly sharing government data | 11 | 12 |
| TB4. Lack of skills to maintain OGD | employees in your municipality lack the skills required for maintaining openly shared government data | 10 | 10 |
| TB5. Reluctancy to change organisational processes | your municipality is reluctant to radically change the organizational processes that would enable openly sharing government data | 10 | 10 |
| Usage barriers | | 12 | 113 |
| Other usage barriers | | 11 | 13 |
| UB1. Inappropriate quality | an inappropriate quality level of your municipality's data | 12 | 13 |
| UB2. Preparation of data complicated | a complicated process to prepare data for sharing | 12 | 12 |
| UB3. Make data reusable complicated | complicated process to make your municipality's data reusable by others | 10 | 12 |
| UB4. Publishing in OD portal complicated | complicated process to publish your municipality's data on an open data portal | 12 | 13 |
| UB5. Publishing of OD unclear | unclear process of publishing your municipality's data on an open data portal | 10 | 11 |
| UB6. Limited functionality of OD portal | limited functionality of open data portals | 11 | 14 |
| UB7. Semi-automate process to share OD | no possibility to semi-automate my municipality's process to openly share its data | 12 | 12 |
| UB8. Maintain data once published | the need and a complicated process to maintain data once published | 11 | 13 |
| Value barriers | | 12 | 113 |
| Other value barriers | | 8 | 9 |
| VB1. OGD not valuable for public | openly sharing government data is often not valuable for the public | 12 | 12 |

| Name | Description | Interviews | References |
|---|---|---|---|
| VB2. Datasets not appropriate for reuse | -many open government datasets are not appropriate for re-use | 12 | 12 |
| VB3. Datasets have quality issues | many open government datasets suffer from data quality issues (completeness, accuracy, uniqueness, consistency, etc.) | 11 | 11 |
| VB4. Public gains are lower than costs | the public gains of openly sharing government data are often lower than the costs | 12 | 13 |
| VB5. Org. gains are lower than costs | your organization's gains of openly sharing government data are often lower than the costs | 7 | 9 |
| VB6. Data prep. too resource-consuming | data preparation is too resource-consuming for your organization | 11 | 11 |
| VB7. OGD provides no value to org. | open government data do not provide any value to your organization | 11 | 11 |
| VB8. Org. OGD provides no value to users | open data that your organization can openly share will not provide value to users | 10 | 13 |
| VB9. Spent resources vs benefits to organization | number of resources to be spent to prepare, publish and maintain open government data outweigh the benefit my organization gains from it | 12 | 12 |